\documentclass[aps,prl,twocolumn,showpacs,groupedaddress]{revtex4}

\usepackage{graphicx}
\usepackage{epstopdf}
\usepackage{color}
\usepackage{url}
\usepackage{hyperref}
\begin{document}

%\preprint{}

\title{Narrowband source of transform-limited photon pairs via four-wave mixing in a cold atomic ensemble}

\author{Bharath Srivathsan}
\altaffiliation{Center for Quantum Technologies, National University of
  Singapore, 3 Science Drive 2, Singapore, 117543}
\author{Gurpreet Kaur Gulati}
\altaffiliation{Center for Quantum Technologies, National University of
  Singapore, 3 Science Drive 2, Singapore, 117543}
\author{Chng Mei Yuen Brenda} 
\altaffiliation{Center for Quantum Technologies, National University of
  Singapore, 3 Science Drive 2, Singapore, 117543}
\author{Gleb Maslennikov}
\altaffiliation{Center for Quantum Technologies, National University of
  Singapore, 3 Science Drive 2, Singapore, 117543}
\author{Dzmitry Matsukevich}
\altaffiliation{Department of Physics, National University of Singapore, 2
  Science Drive 3,  Singapore, 117542}
\altaffiliation{Center for Quantum Technologies, National University of
  Singapore, 3 Science Drive 2, Singapore, 117543}
\author{Christian Kurtsiefer}
\altaffiliation{Department of Physics, National University of Singapore, 2
  Science Drive 3,  Singapore, 117542}
\altaffiliation{Center for Quantum Technologies, National University of
  Singapore, 3 Science Drive 2, Singapore, 117543}

\begin{abstract}
We observe narrowband pairs of time-correlated photons of wavelengths 776\,nm
and 795\,nm from non-degenerate four-wave mixing
in a laser-cooled atomic ensemble of $^{87}${Rb} using a cascade decay scheme.
Coupling the photon pairs into single mode fibers, we observe an instantaneous
rate of 7700 pairs per second with silicon avalanche photodetectors, and an
optical bandwidth below 30\,MHz.
Detection events exhibit a strong correlation in time
($g^{(2)}(\tau=0)\approx5800$), and a high coupling efficiency indicated by a
pair-to-single ratio of 23\%. The violation of the Cauchy-Schwarz
inequality by a factor of $8.4\times10^6$ indicates a strong non-classical
correlation between the generated fields, while a Hanbury--Brown--Twiss
experiment in the individual photons reveals their thermal nature.
The comparison between the measured frequency bandwidth and 1/e decay time of 
$g^{(2)}$ indicates a transform limited spectrum of the photon pairs.
The narrow bandwidth and brightness of our source makes it ideal for
interacting with atomic ensembles in quantum communication protocols.

\end{abstract}


\maketitle

Time-correlated photon pairs have been an important resource for a wide range
of quantum optics experiments, ranging from fundamental tests
\cite{aspect:82,CHSH1969,Fry_1976,weihs:98} to applications in quantum information
\cite{Nielsen:2004,Bouwmeester2000,ekert91}. Most of these
applications, however, are based upon manipulation or detection of photons
only. More complex quantum information tasks require interfacing of
photons to other physical systems, like atoms, molecules, or color centers.
A typical example is a quantum network \cite{kimble2008}, where
information is stored or processed in single ions \cite{blatt_2013},
atoms in a cavity \cite{ritter_2012,Wilk27072007,cirac_1997,luo_1997}, or in
an ensemble of atoms \cite{DLCZ:2001,Kuzmich:2003,vanderwal:2003,thompson2006}.

So far, most of the photon pair sources based on spontaneous
parametric down conversion in $\chi^{(2)}$ nonlinear optical crystals exhibit
a relatively wide optical bandwidth ranging from 0.1 to 2\,THz
\cite{kurtsiefer:01,wong:06}. This makes it
difficult to interact with atom-like physical systems, since their optical
transitions usually have a lifetime-limited bandwidth on the order of several
MHz. Therefore, various filtering techniques have been employed to reduce the
bandwidth of parametric fluorescence light. In addition, the parametric
conversion bandwidth may be redistributed within the resonance comb of an
optical cavity \cite{Wolfgramm:08,Mitchell:2009,Cere:09}. A recent extreme
example uses a ring cavity formed by the nonlinear optical medium itself
\cite{Fortsch:12}.

An alternative approach to this problem is based on four wave mixing (FWM) in
an atomic vapor, which resembles the early approaches for entangled photon
pair preparation via an atomic cascade decay \cite{aspect:82}. In
comparison with atomic beam experiments, which had only a very small
number of atoms participating in the excitation and decay process at any one
time, a cloud of atoms provides a translational symmetry of the nonlinear
medium. This leads to momentum conservation or phase matching for the conversion
process similar to nonlinear optical interaction in suitable crystalline
materials. Momentum conservation in turn allows for a simple
collection of the converted light into optical fibers, which leads to a
relatively high heralding efficiency of one photon. Correlated photon pairs
generated  by FWM via cascade decay in a hot $^{85}\text{Rb}$ atomic ensemble
have been observed  \cite{Willis:11,Ding:12}, with an optical linewidth of
350\,MHz and 450\,MHz, respectively.

In this paper, we report on spontaneous parametric conversion via FWM in a
cold cloud of atoms provided by a Magneto-Optical Trap (MOT), similar to previous work by
Channeliere \textit{et al.} \cite{Chaneliere:2006}. By doing so the Doppler
broadening due to atomic motion is greatly reduced, leading to a bandwidth of 
the collected fluorescence of the cascade decay that is comparable to the
natural atomic line width.

We characterize the temporal properties of the generated photon pairs via a
cross-correlation measurement, and the photon statistics of the signal and
idler photons from a Hanbury--Brown--Twiss experiment. We also determine
the spectral properties of the generated idler photons directly with a
scanning Fabry-Perot cavity.

\begin{figure}
  \begin{center}
    \includegraphics[width=\columnwidth]{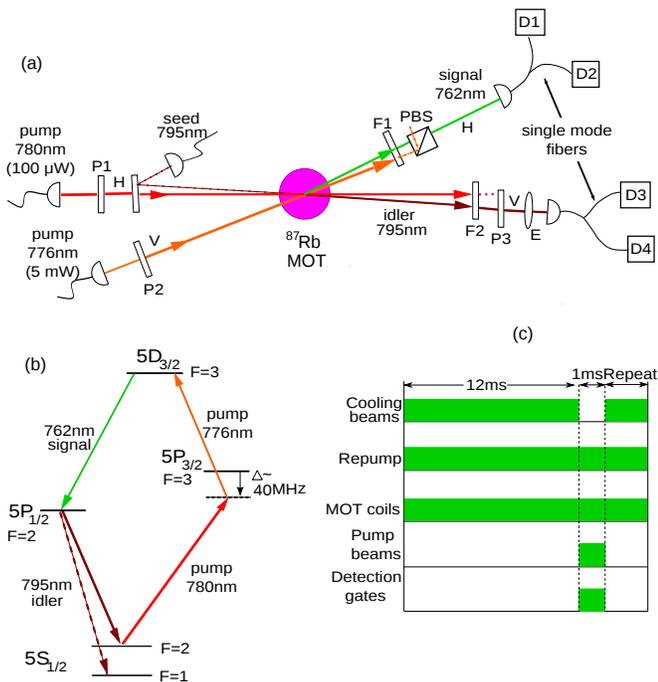}
    \caption{\label{fig:combined_setup_levelscheme.eps}(a) Schematic of the
      experimental set up, with P1-P3: polarization filters, F1, F2:
      interference filters, E: solid etalon, D1-D4: avalanche
      photodetectors. A $795\,\text{nm}$ seed beam is used to determine the phase-matched direction of coherent $762\,\text{nm}$ emission. 
(b) Cascade level configuration in 
      $^{87}\text{Rb}$. (c) Timing sequence for one experimental cycle.  
   }
  \end{center}
\end{figure}

The experimental setup is shown in
Figure~\ref{fig:combined_setup_levelscheme.eps}(a).  
An ensemble of $^{87}\text{Rb}$ atoms is trapped and cooled with a MOT formed
by laser beams red detuned by 24\,MHz from the
$5\text{S}_{1/2},F=2\rightarrow 5\text{P}_{3/2},F=3$
transition, with a diameter of $\approx$ 15\,mm and an optical power of 45\,mW per beam.
An additional laser tuned to the ${5\text{S}_{1/2},F=1\rightarrow
  5\text{P}_{3/2},F=2}$ transition optically pumps the atoms back into the
$5\text{S}_{1/2},F=2$ level. % (10 mW).
With an axial quadrupole field gradient of 0.3\,Tm$^{-1}$ %30Gauss/cm 
we obtain an atomic cloud with a measured optical density (OD$_m$) of $\approx32$,
as determined here and in the following by a fit of the spectral transmission profile of 
a focused probe beam (waist  125\,$\mu$m) around the $5\text{S}_{1/2},\,F=2$
$\rightarrow$  $5\text{P}_{3/2},\,F=3$ transition \cite{fox2006quantum}.

To generate the correlated photon pairs, the MOT is turned off,
and the atoms are excited to the $5\text{D}_{3/2},F=3$ level (see
Fig.~\ref{fig:combined_setup_levelscheme.eps}(b)) by two orthogonally linearly
polarized pump beams (780 nm and 776 nm) intersecting
at an angle of 0.5$^\circ$ in the cold atomic cloud. 
The 780\,nm pump beam is red detuned by 40\,MHz from
the intermediate level $5\text{P}_{3/2},\,F=3$, since its population would
result in a decay back to the initial state.

Experimental periods of 1\,ms for photon pair generation are interleaved with
periods of 12\,ms with the MOT turned on to replenish and cool the atomic
cloud. This duty cycle was experimentally found to lead to the largest
optical density (see Fig.~\ref{fig:combined_setup_levelscheme.eps}(c)).

Photon pairs from a cascade decay of atoms in the excited
$5\text{D}_{3/2},F=3$ level via $5\text{P}_{1/2},F=2$ back into
$5\text{S}_{1/2},F=2$ emerge into well-defined directions determined by
momentum conservation of the four participant modes. Signal and idler photons
generated by parametric conversion are separated from residual  
pump light by interference filters F1, F2 with a bandwidth of 3\,nm full width
at half maximum (FWHM) and a peak transmission of 96\%.
Uncorrelated photons are further removed from signal
and idler modes by a polarizing beam splitter PBS
and polarizer P3, where polarizations of pump and target modes are
chosen to maximize the product of the Clebsch-Gordan coefficients, and
thereby the effective nonlinearity~\cite{Jenkins:07}.
A temperature-tuned solid fused-silica etalon E (linewidth 375\,MHz FWHM, peak
transmission 86\%) in the idler arm is used to remove uncorrelated photons
from a decay to the $5\text{S}_ {1/2},\,F=1$ level. Parametric fluorescence is then coupled into single mode fibers with aspheric lenses. The
effective waists of
the collection modes at the location of the cold cloud were determined to be
0.4\,mm and 0.5\,mm for signal and idler by back-propagating light through the
fibers and couplers. In an initial alignment step,
a seed light at 795\,nm is injected into the idler mode, and coupled into
a single mode fiber with an efficiency of 80\%. The corresponding signal mode
is coupled into the other single mode fiber
with an efficiency of 70\%.

The photons are detected with Silicon avalanche photodetectors (APDs) D1-D4,
(estimated quantum efficiencies of $\approx 40\%$, dark count rates 40 to
150\,s$^{-1}$), and their detection time recorded with a timestamp unit. The
combined timing uncertainty of the detectors and timestamping unit is about
0.6\,ns.

\begin{figure}[]
  \begin{center}
    \includegraphics[width=\columnwidth]{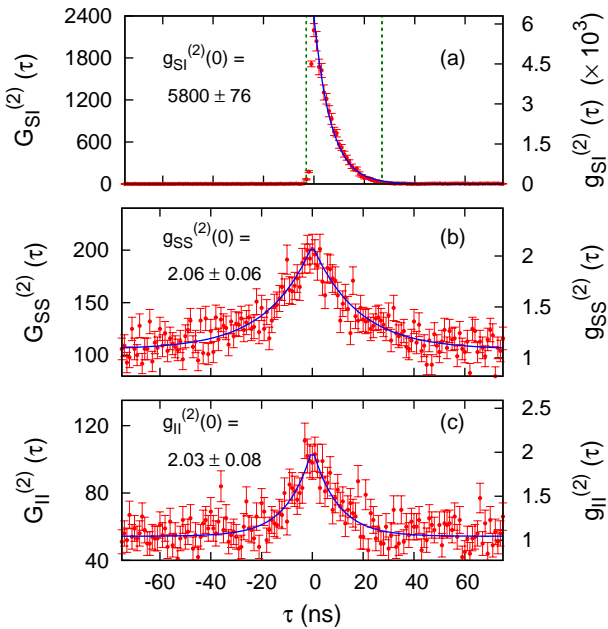}
    \caption{\label{fig:g2}(a) Histogram of coincidence events
      $G_{\text{SI}}^{(2)}(\tau)$ for a total integration time of $T=47$\,s as a
      function of time delay $\tau$ between the detection of idler and signal
      photons sampled into $\Delta\tau=1$\,ns wide
      time bins, and its normalized version $g_{\text{SI}}^{(2)}$ according to
      Eq.~(\ref{eq:g2}). The solid line is a fit to the model $g_{\text{SI}}^{(2)}
      (\tau) = B+A \times \exp(-\tau/\tau_0)$, where $B=1.20\pm0.07$ is the
      mean $g_{\text{SI}}^{(2)} (\tau)$ for $\tau$ from 125\,ns to 1$\mu$s,
      resulting in $A=5800\pm76$ and $\tau_0=6.7\pm0.2$\,ns. (b) Time resolved
      coincidence histogram $G_{\text{SS}}^{(2)}(\tau)$ and its normalized version in
      a Hanbury-Brown--Twiss experiment on signal photons (detectors D1, D2)
      for $T=76.3$\,s. The solid line shows a fit to
      the model $g_{\text{SS}}^{(2)}(\tau)=C\times (1+D\times \exp(-|\tau|/\tau_0))$,
      resulting in $C=1.08\pm0.1$, $D=0.93\pm0.06$ and $\tau_0
      =17.8\pm1.4$\,ns. (c) Same as (b) for idler 
      photons detected on D3 and D4 for $T=247.3$\,s, leading to fit parameters
      $C=1.04\pm0.08$, $D=0.96\pm0.08$, and $\tau_{0} = 9.9\pm1.2$\,ns.
%**991 sec for idler and 3200 for signal**
      For all plots, the atomic cloud has an optical density OD$_{m}\approx32$.
    }
  \end{center}
\end{figure}

The histogram of coincidence events $G^{(2)}_{\text{SI}}(\tau)$ as a function
of time delay $\tau$ between the detection of signal and idler photons
sampled into time bins of width $\Delta\tau$ = 1\,ns is shown in
Fig.~\ref{fig:g2}(a). The normalized cross-correlation is defined as 
\begin{equation} \label{eq:g2}
  g^{(2)}_{\text{SI}}(\tau)\,= \frac{G^{(2)}_{\text{SI}}(\tau)}{r_{\text{I}}
    \, r_{\text{S}}  \, \Delta\tau \, T} \quad,
\end{equation}
where $r_{\text{I}}$=2600\,s$^{-1}$ and $r_{\text{S}}$=3100\,s$^{-1}$ are the
idler and signal photons count rates, and $T$ is the integral time when the
pump beams are on, i.e., 1/13 of the total measurement time
(Fig.~\ref{fig:combined_setup_levelscheme.eps}(b)). The peak at
$g^{(2)}_{\text{SI}}(0)$ of 5800$\pm$76 indicates a strong temporal
correlation. We observe $g^{(2)}_{\text{SI}}(\tau)=1.20\pm0.07$ at a
time delay of 125\,ns to 1$\mu$s, 
with a low decrease to $g^{(2)}_{\text{SI}}(\tau)=1$ at $\tau\approx100\,\mu$s.
The measured $1/e$ decay time for heralded idler photons from the fit
is 6.7$\pm$0.2\,ns, which is lower than the single atom
spontaneous decay time of 27\,ns from $5\text{P}_{1/2},F=2$
level. This is due to the superradiance effect in an optically thick atomic
ensemble ~\cite{Dicke:1954,Rehler:1971}.

A total photon pair detection rate $r_{\text{P}}$ of this source can be derived from
the measured $G^{(2)}(\tau)$ by integrating over a coincidence time window
$\tau_c$, 
$r_{\text{P}}=\frac{1}{T} \sum_{\tau=0}^{\tau_{c}} \! G^{(2)}_{\text{SI}}(\tau)$.  For
$\tau_{c}$ = 30\,ns (vertical lines in Fig.~\ref{fig:g2}(a)), almost all
pairs are captured. Under optimal experimental conditions with pump powers of
5\,mW for 776\,nm, 100\,$\mu$W for 780\,nm, and a detuning
${\Delta}\approx40$\,MHz from the intermediate level
we obtain $r_{\text{P}}=400$\,s$^{-1}$ during the parametric conversion
interval. Under these conditions, we find a signal heralding efficiency
$\eta_{\text{S}} = r_{\text{P}}/r_{\text{S}}= 14.9\%$,
and an idler heralding efficiency $\eta_{\text{I}} = r_{\text{P}}/r_{\text{I}}= 23\%$.
By increasing the 776\,nm pump power to 14\,mW and for a detuning
$\Delta\approx 20$\,MHz from the intermediate level, the (instantaneous) pair rate
increases to $r_{\text{P}}=7700$\,s$^{-1}$, with $g_{\text{SI}}^{(2)}=54\pm7$.
This corresponds to an average detected pair rate of $592$\,s$^{-1}$ 
including the time during which the pump beams are off.
All efficiencies and photon count rates reported are uncorrected for losses
due to non-unit detector efficiency, filtering efficiency and fiber coupling
efficiency.
Correcting for the detector efficiency on both signal
and idler modes, we infer average  and instantaneous rates of usable
photon pairs coupled into the single mode fibers of about 3700 s$^{-1}$
and 48000\,s$^{-1}$, respectively.

While it is well-known that light in each of the modes in parametric
fluorescence should exhibit thermal photon statistics~\cite{mandel_wolf},
the coherence time of most photon pair sources is too short to be directly
observable in an experiment. Due to the long coherence time of the source
presented here, we are able to carry out a direct Hanbury-Brown--Twiss
experiment. The photon counting statistics of signal photons distributed by
a fiber beam splitter onto detectors D1 and D2 (Fig.~\ref{fig:combined_setup_levelscheme.eps}(a)) is shown in Fig.~\ref{fig:g2}(b).
The normalized $g^{(2)}_{\text{SS}}(\tau=0)=2.06 \, \pm \, 0.06$ 
is compatible with $g^{(2)}(0)=2$ of an ideal single mode thermal state within
the statistical uncertainty \cite{barnett2003methods}.
From a similar experiment performed on the idler photons, we also observe thermal statistics ($g^{(2)}_{\text{II}}(0)=2.03\pm0.08$).
Without the solid etalon, the idler photons
coupled into the single mode fiber are of two different frequencies,
thus $g^{(2)}_{\text{II}}(0)<2$ is expected and indeed observed ($1.69\pm0.02$). 

The Cauchy-Schwarz inequality bounds the intensity correlation $g^{(2)}$ between two independent classical fields \cite{Walls:2008,Reid1986}:
\begin{equation} \label{eq:cauchy}
  R = 
\frac{[g^{(2)}_{IS}(\tau)]^{2}}
  {g^{(2)}_{II}(0) \cdot g^{(2)}_{SS}(0)} \leq 1
\end{equation}
This inequality between the signal and idler fields in our
experiment is violated by a factor $R\,=\,(8.4 \pm 0.2) \times 10^6$ at $\tau
=0$ which shows that our source exhibits statistics unexplainable by classical electromagnetic field theory. 
Our violation factor 
strongly exceeds the values reported from similar experiments by Du \textit{et
  al.}~($R=11600$, \cite{Du2008}) and Willis \textit{et al.} ($R=495$,
\cite{Willis:11}). We attribute this to lower background counts 
as compared to what has been observed with gas cells.

\begin{figure}[]
  \begin{center}
    \includegraphics[width=\columnwidth]{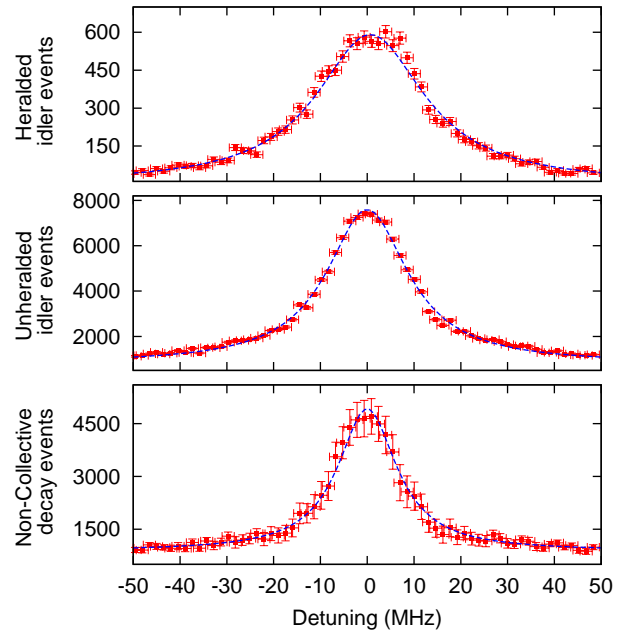}
    \caption{\label{fig:linewidth}(a) Spectral profile of idler
      photons, heralded by the detection of signal photons with an atomic
      cloud OD$_{m}\approx32$. The
      frequency uncertainty is due to the uncertainty in voltage driving
      the cavity piezo. The line shows a fit to a model of Lorentzian-shaped
      photon spectrum, convoluted with the cavity transmission spectrum.
      The fit gives a bandwidth of 24.7$\pm$1.4\,MHz (FWHM). (b) Same as (a),
      but without heralding. The resulting bandwidth from the
      fit is 18.3$\pm$1.3\,MHz (FWHM). 
      (c) Inferred idler spectrum from a two step (non-superradiant)
      decay with 12.4$\pm$1.4\,MHz (FWHM) bandwidth from a fit.
    }
  \end{center}
\end{figure}

An indirect assessment of the bandwidth of the photons can be obtained
from the measured $g^{(2)}(\tau)$, since it is related to the Fourier
transform of the spectral distribution. Assuming a transform-limited spectrum,
we would infer a bandwidth of $\Delta \nu = 1/(2\pi\tau_0)=23.8\pm0.7$\,MHz
(FWHM) for the heralded idler photons (Fig.~\ref{fig:g2}(a)). 

A direct optical bandwidth measurement of idler photons was carried out
with a scanning Fabry-Perot cavity (linewidth 2.8\,MHz FWHM, measured by a
ring-down experiment \cite{Rempe1992}),
tuned $\pm$50\,MHz across the $5\text{P}_{1/2},\,F=2$ $\rightarrow$
$5\text{S}_{1/2},\,F=2$ transition.
To minimize frequency drift the cavity is temperature stabilized to within
10\,mK, and kept in vacuum ($6\times10^{-6}$\,mbar). The central transmission
frequency is periodically recalibrated via a reference laser locked to the
aforementioned atomic transition at 795\,nm.
The results of this measurement (for OD$_{m}\approx32$ of the atomic cloud) is
shown in Fig.~\ref{fig:linewidth}. A fit of the obtained spectrum to a
Lorentzian line shape widened by the cavity transfer function 
leads to a bandwidth of 24.7$\pm$1.4\,MHz (FWHM) for the idler photons, if
they are heralded by a signal photon (see trace (a)). This is comparable with
the bandwidth inferred from the correlation function, indicating that the
photons are indeed transform-limited. 

However, the
observed spectrum of all light in the idler mode (i.e., the unheralded
ensemble) shows a narrower bandwidth of $18.3\pm1.3$\,MHz (FWHM).
This may be explained by incoherent two step decay
(non-collective) contributions to light emitted via the collectively enhanced
decay collected in phase-matched directions. The optical bandwidth of light
from the collective decay contribution should increase with the atom number $N$
due to an enhanced cascade decay rate, while the bandwidth of light
from the two step contribution should remain the same.

\begin{figure}[h]
  \begin{center}
    \includegraphics[width=\columnwidth]{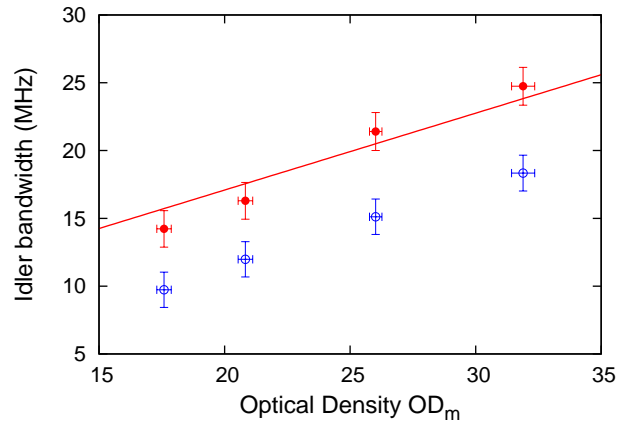}
    \caption{\label{fig:od_linewidth} Bandwidth (FWHM) of heralded idler
      photons (pairs) for different cloud optical densities (OD$_{m}$) (filled
      circles). The line shows the theoretical model according to
      \cite{Walther:2009,Jen:2012}. Open circles indicate the
      bandwidth (FWHM) of unheralded idler light.} 
  \end{center}
\end{figure}

The observed bandwidth $\Gamma$ of idler photons (heralded and unheralded) for
different atomic densities is shown in Fig.~\ref{fig:od_linewidth}, and
increases as expected with the OD$_m$, both for the heralded and unheralded
photon spectrum.

According to \cite{Walther:2009,Jen:2012}, the variation of the emitted
bandwidth $\Gamma$ due to collectively enhanced decay can be modeled with the
relation $\Gamma=\Gamma_0 (1+\mu N)$, with the natural
line width $\Gamma_0=2\pi\times5.8$\,MHz of the $5\text{P}_{1/2},\, F=2$
$\rightarrow$ $5\text{S}_{1/2},\, F=2$ transition~\cite{Steck_2010}, the atom
number $N$ and a geometry factor $\mu$. We also find a linear increase of
$\Gamma$ with OD$_m$ compatible with this model, since $\mu N$ is proportional to
our measured OD$_m$; the solid line in Fig.~\ref{fig:od_linewidth} shows a fit
with the proportionality factor between $\mu N$ and OD$_m$.

Assuming that the incoherent contribution does not significantly
contribute to detected pairs at small numerical apertures for collection, we
can infer its spectrum by subtracting the heralded idler spectrum from the
unheralded idler spectrum after correction for losses in filters (11\%),
optical elements (7\%), inefficient photodetectors (60\%), polarization
filters (12\%), and fiber coupling (30\%). The resulting spectrum for
OD$_m\approx32$ is shown in Fig.~\ref{fig:linewidth}(c), with a width of
12.4$\pm$1.4\,MHz FWHM. This exceeds the natural line width expected
for the incoherent two step decay, probably due to
self-absorption in the atomic cloud.

In summary, the photon pair source presented in this paper exhibits a high
heralding efficiency, is spectrally bright, and shows a narrow optical
bandwidth for signal and idler photons. This narrow Fourier-limited bandwidth
and the
wavelength match with transitions in $^{87}$Rb, a common workhorse for quantum
memories,  makes our source a prime candidate for heralded interaction with single atom systems, 
and quantum memories based on atomic ensembles.
The high normalized cross correlation $g_{\text{SI}}^{(2)}$ value clearly indicates
the non-classical nature of the photon pairs, and a low background rate. 
We also demonstrate the thermal statistics of the signal and idler photons
from a direct autocorrelation measurement. Beyond correlated photon pair
preparation, this scheme can also provide polarization entangled photons by an
appropriate choice of pump polarization~\cite{Chaneliere:2006, Willis:11},
which can used to implement entanglement swapping and other quantum
communication protocols with single atoms 
\cite{Syed:2011,Tey:2008}, ions \cite{eschner_2011}, or atomic
ensembles. Furthermore, the long coherence time of our idler photon heralded
by the `click' detection of the signal photon by an APD enables electric field
quadrature measurements of the idler photon by homodyne detection using
currently available fast photodetectors~\cite{gerber:2009,macrae:2012}.

%\section{Acknowledgment}
We acknowledge the support of this work by the National Research Foundation \&
Ministry of Education in Singapore.
%\bibliographystyle{plain}
% \bibliography{fwm,QIS}

\end{document}